\documentstyle[12pt,aps]{revtex}
\begin{document}

\title{
Chiral symmetry and mixing of axial and vector correlators in matter\bigskip}
\author{ Boris Krippa}
\address{ Institute 
for Nuclear Research of the Russian Academy of Sciences, Moscow Region 117312,
Russia.\\}
\maketitle
\bigskip
\bigskip

\begin{abstract}
The effect of mixing of the vector and axial vector correlation
functions in the nuclear medium arising from the interaction
of nuclear pions with corresponding interpolating currents
is considered. It is shown that the mass difference between $\rho$
and $a_1$ meson gets smaller with the increase of the nuclear density
reflecting the phenomena of partial restoration of chiral symmetry
whereas the absolute values of meson masses may both decrease and increase
in nuclear medium depending on the model used for the phenomenological
spectral density.  
\end{abstract}

During the last few years the behaivior of vector mesons in the dence and/or
hot matter attracted much attention. Among them $\rho$-meson is of particular
interest as, due to rather significant width, $\rho$-meson decays mostly in 
nuclear interior and dilepton invariant spectrum should thus carry the direct
information on the $\rho$-meson properties in the nuclear medium \cite{pi}.
From the theoretical point of view the situation is far from clear.
According to Brown and Rho \cite{br} the rho-meson mass should decrease
in baryon matter like the chiral order parameter (BR-scaling). Indeed, several
 calculations in the framework of QCD sum rules \cite{hl,AK} , devoted to 
the study of density
 dependence of the vector meson masses, suggest significant drop of the
 $\rho$-meson mass with density (about 20$\%$ at the normal nuclear density).
This is the remarkable fact
since the decrease of the rho-meson mass in the nuclear matter may indicate
the partial restoration of chiral symmetry. In
 \cite{hl} nuclear matter was assumed to be a system of nucleons with finite
density and the phenomenological part of the QCD sum rules, represented by
some spectral density which was taken to be
a delta function plus a continuum, whereas in \cite{AK} this spectral
density was treated using the vector dominance model that was extended to 
include the delta-hole polarization initiated by pions produced by the
$\rho$ meson decay in nuclear matter. 
However, in the very recent paper
\cite{kw} it was shown that QCD sum rules analysis results in small
change of the $\rho$-meson mass provided the $\rho$N scattering amplitude
is properly taken into account. Moreover, some effective models
\cite{R93} also lead only to marginal changes of the $\rho$-meson mass in
the matter.

Unfortunately, chiral symmetry alone cannot predict the actual behavior
of the in-medium $\rho$-meson mass. Instead, chiral symmetry dictates the
correlators of the chiral partners to be identical in a chirally symmetric
world. In our case the vector and axial correlation functions should become
equal in the chirally restored phase. Consequently, the vector and axial
correlators get mixed when the chiral symmetry is partially restored.
Such kind of admixture was found in ref. \cite{ei} in the case of finite 
temperature. In this paper we address the same issue to the system with finite
baryon density.

The axial vector correlator includes the contributions from pion and $A_1$
meson. The pionic part of the axial correlator can be determined rather well
at least in the leading order in density, since pion mass remains close to
it vacuum value even for the normal nuclear density and pion decay constant
$F_\pi$ scales with density like $\overline qq$ condensate, in-medium
behaviour of  which can be determined by model-independent way \cite{dl}
in the linear in nuclear density approximation.  The 
contribution  in the axial correlator which is due to $A_1$ meson, causes
more problems. Unfortunately,
there are neither theoretical calculations nor attemps to determine the
in-medium properties of $A_1$ experimentally. Therefore, to study the 
vector axial mixing one needs to assume some in-medium behaviour of the
unmixed correlators. One notes that, whereas the effect of the mixing is 
inevitably present as long as chiral symmetry gets restored, one 
seems impossible to determine quantitatively the in-medium masses of the
chiral partners caused by too large uncertainties already on the level of
the unmixed vector and axial correlators. However one can determine, at least 
qualitatively, the density dependence of the 
$\rho$-$A_1$ mass difference, since this is a parameter
 characterizing the spontaneous breaking of chiral symmetry.

In-medium behaviour of hadrons can be studied theoretically by calculating the
two-point correlator of the corresponding current. In the particalar case
of the $\rho$-meson the correlator is given by   
\begin{equation}
\Pi_{\mu\nu}(p)=
i\!\int d^4x e^{ip\cdot x}\langle \Psi|T\{V_{\mu}(x)V_{\nu}(0)\}
|\Psi\rangle.
\label{correl}
\end{equation}
where $V_{\mu}$ is the isovector current and the matrix element is taken
over the ground states of the system with finite baryon density $\rho$.
In the case of the noninteracting nucleons the expression above is identical
to that, calculated by Hatsuda et.al \cite{hl}. However, when nuclear pions
are explicitely included, the direct interaction of those pions with the 
vector currents may induce the mixing of the vector and axial correlators.
Admixture of the opposite parity channel arises when two pions interact with
vector currents at points 0 and/or $x$. The piece of the correlator where
the vector current interacts directly with the nuclear pion that is emitted
and then absorbed by the nuclear matter can be written in the form
\begin{equation}
\sum_{a,b}\int {d^3\hbox{\bf k}\over 2\omega_k}
{d^3\hbox{\bf k}'\over 2\omega_{k'}}\langle\Psi|a^{a\dagger}(\hbox{\bf k})
a^b(\hbox{\bf k}')|\Psi\rangle
\,i\!\int d^4x e^{ip\cdot x}\langle \Psi\,\pi^a(\hbox{\bf k})|T\{V_{\mu}(x)
V^{\mu}(0)\}|\Psi\,\pi^b(\hbox{\bf k}')\rangle,
\label{piinout}
\end{equation}
In the nuclear matter the pions are virtual so one needs to consider all
possible time orderings where two pions are either emitted or absorbed by
the matter. The corresponding pieces of the correlator can be represented
by the expressions similar to Eq.(\ref{piinout}) but with two forward - or
backward going pions. One has, respectively
\begin{equation}
{1\over 2}\sum_{a,b}\int {d^3\hbox{\bf k}\over 2\omega_k}
{d^3\hbox{\bf k}'\over 2\omega_{k'}}\langle\Psi|a^a(\hbox{\bf k})
a^b(\hbox{\bf k}')|\Psi\rangle
\,i\!\int d^4x e^{ip\cdot x}\langle \Psi|T\{V_{\mu}(x)\,V^{\mu}(0)\}
|\Psi\,\pi^a(\hbox{\bf k})\pi^b(\hbox{\bf k}')\rangle,
\label{twopiin}
\end{equation}
and
\begin{equation}
{1\over 2}\sum_{a,b}\int {d^3\hbox{\bf k}\over 2\omega_k}
{d^3\hbox{\bf k}'\over 2\omega_{k'}}\langle\Psi|a^{a\dagger}(\hbox{\bf k})
a^{b\dagger}(\hbox{\bf k}')
|\Psi\rangle
\,i\!\int d^4x e^{ip\cdot x}\langle \Psi\,\pi^a(\hbox{\bf k})\pi^b(\hbox{\bf 
k}')|
T\{V_{\mu}(x),V^{\mu}(0)\}|\Psi\rangle
\label{twopiout}
\end{equation}
where $\omega_k=\sqrt{\hbox{\bf k}^2+m_\pi^2}$.

We consider the nuclear pions in the chiral limit so that the nuclear wave
functions appearing in the momentum and space integrals are the same 
in the leading chiral order implying that in this order pion absorption 
or/and emission does not change the nuclear wave functions. One notes,
that similar approach \cite{bk} was used earlier to remove the unwanted
pieces forbidden by chiral symmetry in the QCD sum rules for the nucleon mass
in matter. In some sence in ref. \cite{bk} and in the present paper
the same issue, namely the consistency of the QCD sum rules calculations
of hadron properties in the nuclear matter, is addressed. In case of
$\rho$-meson
this amounts to necessity to include into consideration the in-medium 
behaviour  of its chiral partner, $A_1$ meson. In other words, any study
of the $\rho$-meson mass in matter is not fully consistent with chiral
symmetry, unless the in-medium modifications of $A_1$ meson is included.
Using the soft-pion theorem one has
\begin{equation}
i\langle \Psi\,\pi^a(\hbox{\bf k})|T\{V_{\mu}(x),
V^{\mu}(0)\}|\Psi\,\pi^b(\hbox{\bf k}')\rangle
\simeq {-i\over f_\pi^2}\langle \Psi|\left[Q_5^a,
\left[Q_5^b, T\{V_{\mu}(x),V^{\mu}(0)\}\right]\right]|\Psi\rangle,
\label{soft}
\end{equation}
while the matrix elements in both (\ref{twopiin}) and (\ref{twopiout}) also
reduce to the same expression. 
Calculating the double commutator and making use of the usial expansion of 
the pion field in terms of creation and annihilation operators and commutation
relations of current algebra
$\left[Q_5^a, J_\nu^b\right]=i\epsilon^{abc}A^{c}_\nu$
one can get the following expression for $\Pi_V$
\begin{equation}
\Pi_V=\overcirc\Pi_V+2\xi(\overcirc\Pi_V-\overcirc\Pi_A)
\label{PiV}
\end{equation}
and similar expression for the axial correlator $\Pi_A$
\begin{equation}
\Pi_A=\overcirc\Pi_A+2\xi(\overcirc\Pi_A-\overcirc\Pi_V)
\label{PiA}
\end{equation}
Here $\overcirc\Pi_V$($\overcirc\Pi_A$) is the
correlator of the vector (axial) currents
calculated in the approximation of the noninteracting nucleons with finite
density. We defined
\begin{equation}
\xi={\rho\overline{\sigma}_{\pi{\scriptscriptstyle N}}\over f_\pi^2 m_\pi^2}.
\end{equation}
and used the relation
\begin{equation}
\overline{\sigma}_{\pi{\scriptscriptstyle N}}=4\pi^{3} m_\pi^2 <N|\pi^{2}(0)|N>
\end{equation}
where $\overline{\sigma}_{\pi N}\simeq -25$ MeV is the leading nonanalytic term
in the $\sigma_{\pi N}$ corresponding to the contribution of the pion mass term
to the nucleon mass and $\pi(0)$ is the pion field. From the Eqs. (6-7)
one can see that the complete mixing arises at $\rho\simeq 3\rho_0$ but
this estimate
should not be taken too seriously, since both the in-medium condensates
and phenomenological spectral anzats get highly model dependent at such
densities and reliable conclusion cannot be drawn. Moreover, the soft-pion
approximation may no longer be valid near the point of chiral symmetry
restoration. It is worth mentioning that,
even at the normal nuclear density, where 
the change of the two-quark condensate can almost entirely be described
by the model independent, linear in nuclear density term, only the qualitative
conclusions about the in-medium behaviour of axial and vector meson mass 
difference can be made. This is basically due to unknown behaviour of the
correlator of the axial currents in the nuclear medium. We assume that 
density dependence of $\overcirc\Pi_A$ is the same as that for
$\overcirc\Pi_V$. Besides, the higher twists contributions and the
factorization approximation, which is clearly too crude to be used in nuclear
matter, make the OPE for the vector correlator full of uncertainties too.
One notes, that there is no distinction between the longitudinal and transverse
directions in nuclear medium and, as such, the corresponding correlators can
be related to each other $\Pi_T=\omega^2\Pi_L$ and
$3\omega^2\Pi_L=-g^{\mu\nu}\Pi_{\mu\nu}$ so only the longitudinal correlation
function is really needed. Representing the correlators by the standard 
expressions containing pole terms plus continuum contributions, starting
at some effective thresholds and then applying the Borel transform one can 
get the following sum rules
\begin{equation}
F^{2}_{\rho} exp(-M^{2}_{\rho}/M^2)-
{1\over 8\pi^2}M^{2}exp(-S_{V}/M^2)=\overcirc\Pi_V+
2\xi(\overcirc\Pi_V-\overcirc\Pi_A)
\end{equation}
and
\begin{equation}
F^{2}_{A_1} exp(-M^{2}_{A_1}/M^2)+F^{2}_{\pi}-
{1\over 8\pi^2}M^{2}exp(-S_{A_1}/M^2)=\overcirc\Pi_{A_1}+
2\xi(\overcirc\Pi_{A_1}-\overcirc\Pi_V)
\end{equation}
The unmixed correlators are given by
\begin{equation}
\overcirc\Pi_V=\overcirc F^{2}_{\rho} exp(-\overcirc M^{2}_{\rho}/M^2)-
{1\over 8\pi^2}M^{2}exp(-S_{V}/M^2)
\end{equation}
\begin{equation}
\overcirc\Pi_{A_1}=\overcirc F^{2}_{A_1}exp(-\overcirc M^{2}_{A_1}/M^2)
+\overcirc F^{2}_{\pi}-
{1\over 8\pi^2}M^{2}exp(-S_{A_1}/M^2)
\end{equation}
Here $\overcirc F$ and $\overcirc M$ is the density dependent
residue and mass of the
corresponding hadron calculated without the pion corrections. In case of 
the $\rho$-meson we used the results obtained in Ref.\cite {jl} for
$\overcirc F$ and $\overcirc M$. In principle  effective thresholds are
also different with and without pionic corrections. However, the numerical
effect is small and, therefore, we keep effective thresholds the same.
In the practical calculations we used the 
values $S_{A_1}$=2.2GeV and $S_{V}$=1.5GeV.
As we already mentioned, the in-meidium behaviour of
$\overcirc F_{A_1}$ and $\overcirc M_{A_1}$ is completely unknown so we assumed
that they scale like the corresponding parameters of the vector spectral
density,
starting from the values $\overcirc F^{2}_{A_1}=0.021GeV^2$ and
 $\overcirc M_{A_1}=1.26GeV$ at zero density. The density dependence
of $\overcirc F_{\pi}$ can be found from the in-medium analog of the GOR
relation $F^{2}_{\pi}M^{2}_{\pi}=2M_q<q\overline q>_{\rho}$ making use of the
fact that the pion mass remains almost unchanged at least at normal nuclear
density and using the values of $<q\overline q>_{\rho}$ both with 
and without pionic corrections obtained by
Weise et.al \cite{bw}.
 One notes that at $\rho=\rho_{0}$ the value of $F^{2}_{\pi}$ is
quite close
to $\overcirc F^{2}_{\pi}$ since at normal nuclear density
 meson corrections gives only minor
contribution to the quark condensate \cite{bw}. In order to obtain the 
expressions for $M_{A_1}$ and $M_\rho$ one needs to get rid of the 
unknown factors $F_{A_1}$ and $F_\rho$. It can be done by differentiating
the sum rules over the variable $-1/M^2$ and dividing the obtained expression
by the original sum rules. The corresponding masses $M_\rho$, $M_{A_1}$ are
shown in Fig.1 as the functions of the nuclear density. One can see from
Fig.1
that, although the rho mass still decreases with an increase of the nuclear
density, the size of this effect is significantly smaller than in the case
of the unmixed correlator, whereas the $A_1$ meson mass drops considerably
faster resulting in the convergence of the rho and $A_1$ masses. It may
indicate the tendency toward the partial chiral symmetry restoration.
One can also try to use the other model for the phenomenological part of the
vector correlator developed in Ref.\cite{kw} and leading to the almost
unchanged in-medium mass of the rho-meson. As one can see from Fig.2
in this case $M_\rho$ grows with density while $M_{A_1}$ still decreases,
albeit quite slowly. It is worth mentioning that, in spite of the distinctions
in the density dependence for the vector and axial meson masses, obtained
in different models, the mass difference $M_\rho - M_{A_1}$ remains
approximately the same in all cases. However, one needs to stress again
that the above results are obtained under the assumption that
$\overcirc\Pi_V$ and $\overcirc\Pi_A$ scale identically with density.
One could do the complete QCD sum rules calculations for the axial channel
but the uncertainties will be at least as large as those for QCD sum rules 
analysis of the in-medium rho-mass so it seems unlikely that such culculations
will significantly reduce the ambiguities in the value of the ${A_{1}}$
mass in the nuclear matter.     

In this paper we have presented the analysis of the pionic corrections in
the QCD sum rules for the vector and axial vector mesons in the nuclear matter.
 It was found that such corrections result in the effect of mixing of the 
correlators  of the opposite parity, calculated in the approximation when
the nuclear matter was treated as a medium of the noninteracting nucleons
with some finite density. The mass difference between the chiral partners
$\rho$ and $A_{1}$ decreases with the growth of the nuclear density
indicating the tendency towards the partial restoration of chiral symmetry.
However, chiral symmetry alone cannot predict the actual values of the meson
masses in the nuclear medium due to uncertainties both in the in-medium
quark and gluon condensates (especially in the 4-quark condensate) and in
the phenomenological spectral densities. The standard anzats represented
by the sum of pole and continuum seems to be unadequate to get reliable
density dependence of the in-medium masses. 
In order to go beyond the normal nuclear density one needs to consider some 
additional effects. First,the soft-pion approximation may become invalid
if the density is high enough so at some point the finite pion momenta should
be taken into account. Second, at high densities, when the $\rho$-$A_1$ mass
difference may become comparable with the typical energy of the nuclear 
transitions, the exitation of the residual nucleus should be included so that
the effect of mixing of vector and axial vector correlators will be represented
by the sum over $p-h$ states with certain parity to maintain the corresponding
conservation law. Besides, at high densities the direct interactions of the
isovector current with heavy scalar mesons may also give a significant
contribution.

\section*{Acknowledgements}

The author gratefully acknowledge useful discussions with M.Birse, T.Hatsuda
and S.H.Lee.

\newpage
\begin{center}
{\Large FIGURES CAPTIONS}
\end{center}
\vskip1cm
Fig.1  Density dependence of the mass (in GeV) of the $A_1$(upper curve) and
$\rho$ (lower curve) mesons expressed in the units of the normal nuclear density
$\rho_0$ when the pole plus continuum model \cite{hl,jl} is used for  
$\overcirc\Pi_V$ and $\overcirc\Pi_{A_1}$.\\
Fig.2  The same as in Fig.1a but with the use of the model for
$\overcirc\Pi_V$ and $\overcirc\Pi_{A_1}$ suggested in \cite{kw}.
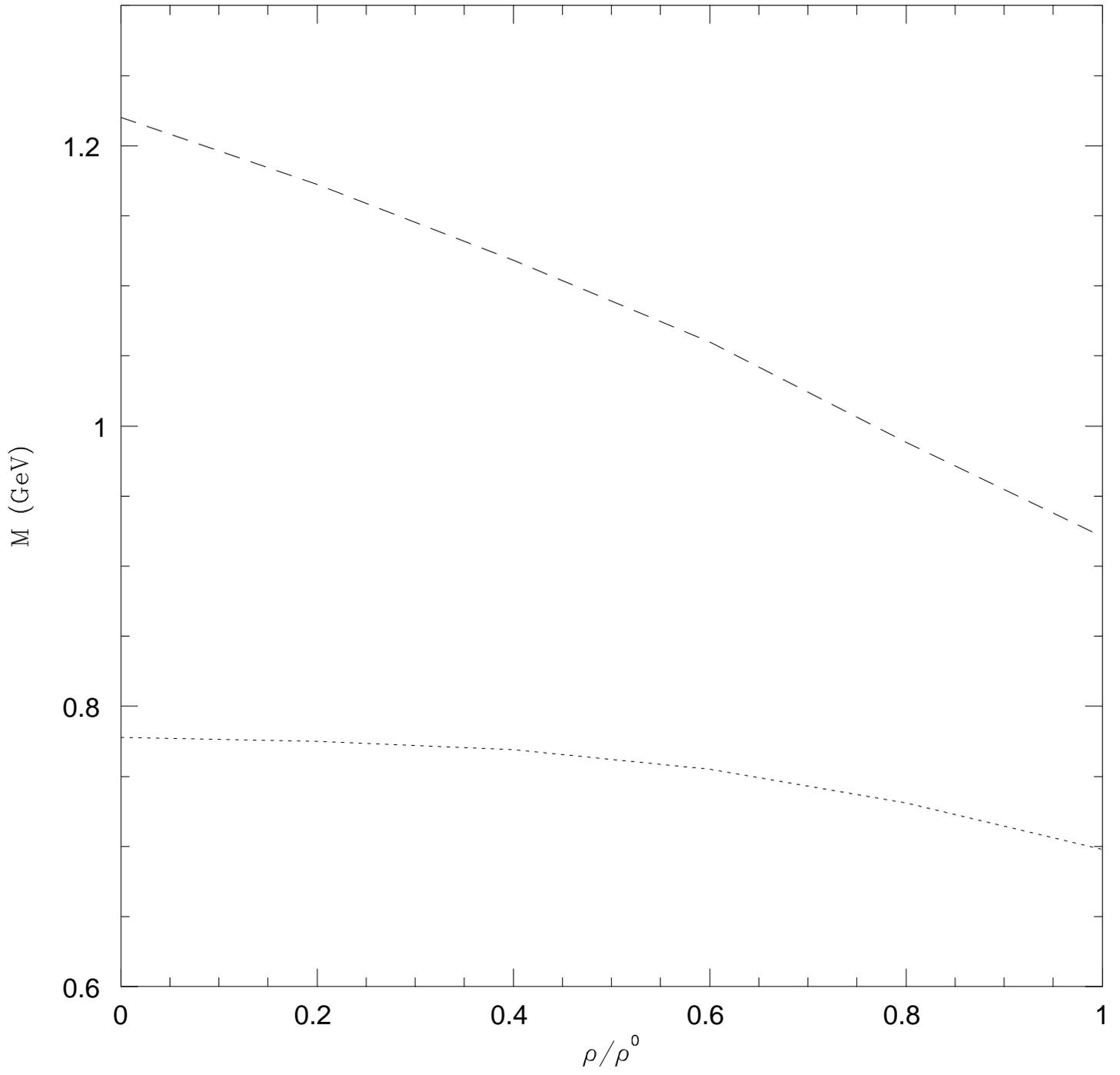
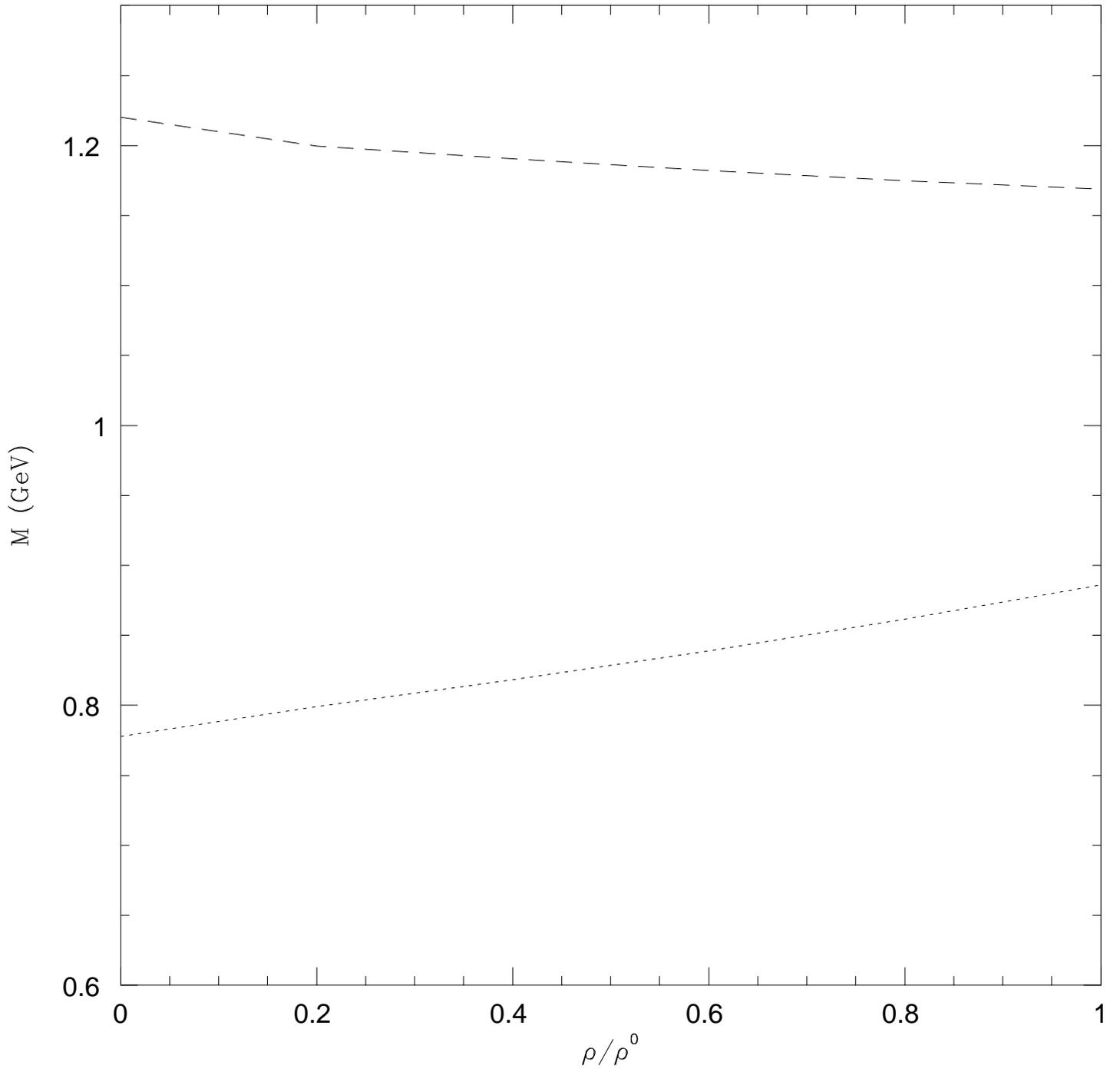
\end{document}